\documentclass[aps,pre,twocolumn,showpacs,superscriptaddress,groupedaddress,10pt]{revtex4-1}  

\usepackage{url} 
\usepackage{graphicx}  
\usepackage{subcaption}
\usepackage{grffile}
\usepackage{dcolumn}   
\usepackage{bm}        
\usepackage{amssymb}   
\usepackage{amsmath}
\usepackage{amsthm}
\usepackage{mathtools}
\usepackage{xcolor}
\usepackage{tikz}	
\usepackage{ulem}
\usepackage[justification=raggedright,singlelinecheck=false]{caption}
\usepackage{subcaption}
\usepackage[ansinew]{inputenc}
\usepackage{calrsfs}
\renewcommand{\emph}[1]{\textit{#1}}

\DeclareMathOperator*{\Tr}{Tr}

\begin{document}



\title{Ensemble-dependent effect of disorder on multicritical points in a system with long-range interactions}

\date{\today}

\author{Nir Schreiber}
\email[]{nir.schreiber@gmail.com}
\affiliation{Department of Mathematics, Bar Ilan University, Ramat Gan, Israel 5290002}
\author{Reuven Cohen}
\affiliation{Department of Mathematics, Bar Ilan University, Ramat Gan, Israel 5290002}

\begin{abstract}
We introduce two variants of the Blume--Emery--Griffiths model with
long-range, mean-field-like interactions and quenched disorder. Although
the two variants are described by different Hamiltonians, they yield the
same free energy. Moreover, both Hamiltonians lead to the same average
energy and hence to the same microcanonical entropy. The canonical and
microcanonical phase boundaries differ markedly, and so does the
multicritical behaviour of the two ensembles. In the microcanonical ensemble, tricritical points appear for any fraction of active spins, whereas in the canonical ensemble they emerge
only when this fraction is sufficiently large. At lower fractions,
canonical critical endpoints arise instead. We also report
the formation of gaps in the entropy as a function of magnetization,
another manifestation of ensemble inequivalence.
\end{abstract}

\maketitle


\section{Introduction}
\label{sec:intro}

Systems with long-range interactions (LRI) are those in which the pair
potential decays at large distance as $U(r)\sim r^{-\alpha}$, with
$0\le\alpha\le d$ in $d$ dimensions~\cite{Campa2009,Campa2014,Dauxois2002,Bouchet2010,Levin2014}. They arise
in self-gravitating systems~\cite{Padmanabhan1990,Chavanis2006}, plasmas
and two-dimensional vortex flows~\cite{Onsager1949,Joyce1973,Miller1990,Robert1992}, wave-particle
systems such as the free-electron laser~\cite{Barre2004FEL}, and in
paradigmatic mean-field models such as the Hamiltonian mean-field (HMF)
model~\cite{Antoni1995}. LRI are also now realized and controlled in
cold atoms coupled to optical cavities~\cite{Ritsch2013}, trapped-ion
crystals~\cite{Britton2012}, Rydberg-atom arrays~\cite{Saffman2010}, and
dipolar quantum gases~\cite{Lahaye2009}.

For a homogeneous system in a hypersphere of radius $R$, the energy per
particle is dominated by $\int^{R} r^{d-\alpha-1}\,dr\sim R^{d-\alpha}$,
for $\alpha<d$, which diverges as $R\to\infty$ \footnote{For $\alpha = d$ the integral 
displays a logarithmic divergence with $R$} . The energy is therefore
non-extensive, i.e., it does not scale with the volume $V\sim R^{d}$.
This can be corrected by rescaling the coupling with the system size, as
in the Kac prescription~\cite{Kac1963,Lebowitz1966}. A more
fundamental difficulty remains, that is, the energy is generally
\textit{non additive}. A system of energy $E$ cannot be split into two
subsystems with energies $E_1$, $E_2$ such that $E=E_1+E_2+o(V)$,
because the coupling between the parts is of the same order as the
energies of the parts. Non-additivity has well-known consequences in the
microcanonical ensemble. For instance, negative specific heat
\cite{LyndenBell1968,Thirring1970}, breaking of ergodicity
\cite{Mukamel2005}, long-lived quasi-stationary states
\cite{Yamaguchi2004}, and, most importantly for what follows,
\textit{inequivalence of ensembles}
\cite{Dauxois2000,Barre2001,Barre2002}. The large-deviation formalism
shows that the canonical and microcanonical ensembles are equivalent
when the microcanonical entropy is concave in the energy (and in the
other conserved or order-parameter-like variables), and may disagree
when it is not~\cite{Ellis2000,Touchette2004,Touchette2009}.

An interesting manifestation of ensemble inequivalence is captured by
multicritical points. 
In a spin system, the tricritical point, at which a second-order line becomes first order
\cite{Lawrie1984}, need not coincide in the two ensembles; the same may
hold for a critical endpoint (CEP), where a critical line terminates on a
first-order line~\cite{FisherUpton1990}. This was demonstrated for the
mean-field Blume--Emery--Griffiths (BEG) model~\cite{BEG1971}, a
spin-one model, introduced for
$^3$He-$^4$He mixtures and with the Hamiltonian
$H=H_I+\Delta\sum_i\sigma_i^2$, where $H_I$ captures the spin-spin interaction, 
$\Delta$ is the crystal field (CF) and $\sigma_i\in\{-1,0,1\}$.
Its microcanonical solution revealed canonical and microcanonical
phase diagrams that differ in the CF-temperature plane, in particular
with different tricritical points~\cite{Barre2001,Barre2002}. The
mean-field BEG model has since been analyzed from the viewpoint of
thermodynamic versus statistical nonequivalence~\cite{Ellis2004} and
through a complete analysis of its ensemble inequivalence
\cite{Hovhannisyan2017}. The rich global phase structure of the model,
including tricritical points, critical endpoints, and their evolution
with the model parameters, is well documented
\cite{Lajzerowicz1975,Furman1977,Hoston1991}.

In all these studies the Hamiltonian is translationally invariant. Real
systems, however, contain quenched disorder, and it is natural to ask
how it interacts with ensemble inequivalence. 
Within the framework of the BEG model, several variants with quenched
disorder have been studied in the mean-field approximation. For instance,
models in which the disorder enters through a random applied magnetic
field or a random CF, each following a bimodal distribution, were explored
using the replica approach~\cite{Salmon2010,Santos2015} or large-deviation
techniques~\cite{SumedhaJana2017,SumedhaMukherjee2020}. Mukherjee and
Sumedha~\cite{MukherjeeSumedha2023} distinguished CEPs from tricritical
points through singularities in the curvature of the phase boundary and
in the coexistence diameter. Benyoussef et al.~\cite{Benyoussef1987}
reported a reentrant portion of the second-order line that meets the
first-order line. They also conjectured a double CEP followed by a
CEP~\cite{SumedhaMukherjee2020}.


The analyses in the works cited in the previous paragraph were done in the canonical ensemble.
To uncover ensemble inequivalence 
where the control parameter $p$ is associated with quenched disorder,
a microcanonical treatment is required. The purpose of the present paper is to fill the remaining gap.
Strictly speaking, 
we address the problem in a minimal setting: the BEG
model on a complete graph with quenched random CF. 
We recover the exact canonical free energy~\cite{Benyoussef1987} and
obtain the microcanonical entropy, for arbitrary $p$. We then compare the resulting
phase diagrams in the CF-temperature plane. An emphasis is put
on the canonical and microcanonical multicritical points and their evolution with $p$.

The rest of the paper is organized as follows. In Sec.~\ref{sec:model},
we introduce our model and present its solution in both ensembles.
Using the thermodynamic functions obtained there, we analyze the
critical behavior of the model in Sec.~\ref{sec:critical}. Concluding
remarks are given in Sec.~\ref{sec:conclusion}.

\section{Model}
\label{sec:model}
We present a model of a system of $N$ interacting spins that realizes two types of randomness, governed by different BEG Hamiltonians. In one realization, the spins are homogeneous and interact with a randomly chosen field. In the other realization, the spins are effectively heterogeneous; that is, a random fraction of the spins interact with a uniform field while the remaining spins do not. Although treated differently in the canonical ensemble, both realizations lead to the same free energy. The two Hamiltonians also yield the same average energy, which is the starting point of the microcanonical solution and leads to a single entropy. Below, we describe the solutions of the model in the canonical and microcanonical ensembles in more detail.

\subsection{Canonical ensemble}
\subsubsection{Random field and homogeneous spins}
\label{sec:RF}
Consider a system governed by the Hamiltonian
\begin{equation}
\label{eq:H}
H = -\frac{J}{2N}\left(\sum_i \sigma_i\right)^2 + \sum_i\Delta_i\sigma_i^2\;,
\end{equation}
where $J>0$ is the ferromagnetic coupling strength (we assume henceforth
for simplicity that $J$ together with Boltzmann's constant $k_B$ are equal to one)
and $\Delta_i/\Delta\sim \mathrm{Ber}(p)$ are {\it{i.i.d}} random variables with $\Delta>0$.

We start with the free energy density in the thermodynamic limit
\begin{equation}
\label{eq:F_can_main}
\beta f  = \lim_{N\to\infty} -\frac{1}{N}\langle\ln Z\rangle\;,
\end{equation}
where $\beta = 1/T$ is the inverse temperature, $Z$ is the partition function 
associated with the Hamiltonian Eq.~\eqref{eq:H} and the average is taken with respect to the disorder. To compute $\langle\ln Z\rangle$, 
we employ the replica trick~\cite{mezard1987spin,dotsenko2001introduction,salmon2009multicritical}, based on the identity
\begin{equation}
\label{eq:replica_id_main}
\ln Z = \lim_{n\to 0}\frac{Z^n-1}{n}\;,
\end{equation}
hence the problem is reduced to averaging over $Z^n$.
It is shown in appendix \ref{AppA} that 
\begin{eqnarray}
\label{eq:free_energy_Ns1}
\beta f  &=& \frac{1}{2}\beta m^2-p\ln\left(2e^{-\beta \Delta}\cosh(\beta m)+1\right)\nonumber\\
&-& (1-p)\ln\left(2\cosh(\beta m)+1\right)\;,
\end{eqnarray} 
where $m$ is the magnetization satisfying
\begin{equation}
\label{eq:mag}
m = \frac{p\sinh(\beta m)}{\cosh(\beta m)+ e^{\beta \Delta}/2} + 
\frac{(1-p)\sinh(\beta m)}{\cosh(\beta m)+1/2}\;.
\end{equation}

\subsubsection{Homogeneous field and random spins}
\label{sec:RS}
In the present section we formulate the problem as a problem 
of choosing at random $N_a\sim \mathrm{Bin}(N,p)$ \textit{active spins}, i.e., spins
that interact with a uniform positive field $\Delta$.
The rest $N-N_a$ spins are inert with respect to the field.
The new Hamiltonian reads
\begin{equation}
\label{eq:H_new}
H_a = -\frac{1}{2N}\left(\sum_i\sigma_i\right)^2 + \Delta\sum_{I}\sigma_{I}^2\;,
\end{equation}
where the index $I$ refers to an \textit{active site}, or a site that hosts an active spin.
The summation runs over all the active sites and satisfies $\sum_I = N_a$.
It is shown in appendix \ref{AppB} that the partition function can take the form 
\begin{equation}
\label{eq:Z_strong_main}
Z_a = \sqrt{\frac{N\beta}{2\pi}} \int_{-\infty}^\infty dx
e^{-N\tilde f_{N_a}(x)}\;,
\end{equation}
where 
\begin{eqnarray}
\label{eq:free_energy_Ns}
\tilde f_{N_a}(x) &=& \frac{1}{2}\beta x^2-\frac{N_a}{N}\ln\left(2e^{-\beta \Delta}\cosh(\beta x)+1\right)\nonumber\\
&-&\frac{N-N_a}{N}\ln\left(2\cosh(\beta x)+1\right)\;.
\end{eqnarray}
Now, by applying the saddle point approximation to Eq.~\eqref{eq:Z_strong_main}
and following~\cite{schreiber2023ensemble}, we can write the free energy density (up to terms $o(1)$) as 
\begin{equation}
\beta f  = \min_x\langle \tilde f_{N_a}(x)\rangle\;,
\end{equation}
which, together with the minimizer, $m$, recovers Eqs.~\eqref{eq:free_energy_Ns1},\eqref{eq:mag}.


\subsection{Microcanonical ensemble}
\label{sec:microcanonical}
\setcounter{subsection}{0}
Let $N_{\pm},N_0$ be the number of spins with values $\pm 1,0$, respectively.
Denote by $N_{a_\pm}$ the number of active spins taking 
the values $\pm 1$.
The total energy, expressed in terms of these occupation numbers, is given by
\begin{equation}
\label{eq:energy0}
E = -(N_+-N_-)^2/2N + \Delta(N_{a_+}+N_{a_-})\;.
\end{equation}
Taking $N_{a_\pm} \approx \langle N_{a_\pm}\rangle = N_\pm\frac{N_a}{N}\approx pN_\pm$, Eq.~\eqref{eq:energy0} can be rewritten, to leading order in $N$, as
\begin{equation}
\label{eq:energy}
E = -(N_+-N_-)^2/2N + p\Delta(N_++N_-)\;.
\end{equation}
It is easy to see that Eq.~\eqref{eq:energy} can be alternatively derived by 
considering the model in section \ref{sec:RF}, i.e., coupling $N$ spins to the single particle fields and replacing $\Delta_i$ in Eq.~\eqref{eq:H} with $\langle \Delta_i\rangle = p\Delta$.

The number of states with energy $E$ is given by 
\begin{equation}
\Omega = \binom{N}{N_+,N_-,N_0}\;.
\end{equation}
To express the entropy density, $s=N^{-1}\ln\Omega$, in terms of 
the parameters of the model, we first divide Eq.~\eqref{eq:energy} by $N$ and write, taking $\epsilon = E/N$,
\begin{equation}
\label{eq:energy_per_spin}
\epsilon = -\frac{1}{2}m^2 + p\Delta q\;,
\end{equation}
where $m = (N_+-N_-)/N$ and $q = (N_+ + N_-)/N$
are the magnetization and quadrupole moments, per spin, respectively.
Next, we introduce the normalized variables 
$\xi_{\pm} = N_{\pm}/N,\ \xi_0 = N_0/N$ satisfying $\xi_+ + \xi_- + \xi_0=1$.
One can then write $\xi_+ = (q+m)/2,\ \xi_- = (q-m)/2$
and obtain, with the aid of Eq.~\eqref{eq:energy_per_spin}, 
\begin{eqnarray}
\label{eq:normalized}
\xi_\pm &=& \frac{2\epsilon + m^2 \pm 2p\Delta m}{4p\Delta}\;,\nonumber\\
\xi_0 &=& -\frac{2\epsilon + m^2 -2p\Delta}{2p\Delta}\;.
\end{eqnarray}
Taking the thermodynamic limit of $s$ and using Eq.~\eqref{eq:normalized}
finally leads to
\begin{equation}
\label{eq:entropy}
s = -\sum_{i\in\{+,-,0\}} \xi_i \ln \xi_i\;,
\end{equation}
which, through the $\xi_i$, is a function of $\epsilon$, $m$, $\Delta$,
and $p$.


\section{Results}
\label{sec:critical}
In the present section we obtain exact expressions for the canonical
and microcanonical continuous transition lines. Exact expressions for
the $p$-dependent CTPs and MTPs are derived. It is shown that, in the canonical ensemble, CEPs exist in a certain range of concentrations. Analysis of the phase diagrams and resulting multicritical 
points is performed.

\subsection{Canonical ensemble}
We start with the canonical ensemble, by expanding the RHS of Eq.~\eqref{eq:mag} in small $m$ 
\begin{equation}
\label{eq:mag_expansion}
m = \left(\frac{2 \beta(1-p)}{3}+\frac{2 \beta p}{e^{\beta \Delta}+2}\right)m +....
\end{equation}
It can be shown that the next nonzero (cubic) coefficient is always negative, thus, Eq.~\eqref{eq:mag_expansion} has a 
nontrivial solution if the linear coefficient is larger than one.
The marginal case where the latter is equal to one determines the second order line
\begin{equation}
\label{eq:can_2nd_strong}
\left(3-2\beta(1-p)\right)\left( e^{\beta\Delta} +2\right) = 6\beta p\;.
\end{equation}

An alternative approach that also probes the
\textit{canonical tricritical point} (CTP) is to write the free energy as a 
Landau normal form around $m=0$,
\begin{equation}
\label{eq:lada_expansion_strong}
f = A m^2+ B m^4 + C m^6 + D m^8 + O(m^{9})\;,
\end{equation}
where the (unimportant) zero order term, corresponding to 
the free energy at $m=0$, is omitted. The higher order coefficients, using the substitution $v=e^{\beta\Delta}$, are given by
\begin{eqnarray}
\label{eq:landau_coeff_strongA}
A &=&
\frac{1}{2} - \frac{\beta(1-p)}{3} - \frac{\beta p}{v+2}\;,\\
\label{eq:landau_coeff_strongB}
B & = &
\frac{\beta^3 \left((1-p)v^2-(7 p-4)v+8 p+4\right)}{36 \left(v+2\right)^2}\;,\\
\label{eq:landau_coeff_strongC}
C & = &-
\frac{\beta^5}{3240 \left(v+2\right)^3}\Big(13 (1-p) v^3+
(78-69p) v^2\nonumber\\
&+&78 (2-5p) v+8 (13 + 59 p)\Big)\;,\\
D &=&
\frac{\beta^7}{60480 \left(v+2\right)^4} 
\Big(41 (1-p)v^4+(328-331 p) v^3\\
&+& (984-624 p)v^2
+(1312 -4876 p) v+ 656 + 5872 p\Big)\;.\nonumber
\end{eqnarray}
To have a second order transition $A$ must change sign at the critical
point while keeping $B> 0$. Indeed, when $A=0$, Eq.~\eqref{eq:can_2nd_strong} is recovered.
Setting $B$ to zero in Eq.~\eqref{eq:landau_coeff_strongB} yields
real solutions $v_1\leq v_2$ iff $p\geq  p_0=\frac{8}{9}=0.8888...\ $.
Hence a second order transition may be exhibited for $p<p_0,\ \forall v$. In other words, in that range of concentrations, a second order line $T_c(\Delta)$ that holds for every positive $\Delta$  may be found.

A CTP $(\Delta^\ast,\beta^\ast)$,
where $\Delta^\ast = \ln (v^\ast)/\beta^\ast$, emerges provided $A=B=0$ and $C>0$.
Setting Eqs.~\eqref{eq:landau_coeff_strongA},\eqref{eq:landau_coeff_strongB} to zero and solving simultaneously for $v,\beta$ give
\begin{eqnarray}
\label{eq:CTP_as_a_fnc_p_Delta}
v^\ast &=&  \frac{7 p-4-3 \sqrt{9 p^2-8p)}}{2 (1-p)}\;,\\
\label{eq:CTP_as_a_fnc_p_beta}
\beta^\ast &=& \frac{3 \left(4-3p-\sqrt{9 p^2-8 p}\right)}{8 (1-p)}\;,
\end{eqnarray}
where we take $v^\ast=v_1$
\footnote{The limit $p\to 1$ does not exist for the other solution of $B=0$, $v_2$, 
with a plus sign before the square root in Eq.~\eqref{eq:CTP_as_a_fnc_p_Delta}. Thus, by continuity, 
this solution is ruled out.}. Indeed, by substituting $v^\ast$ 
in Eq.~\eqref{eq:landau_coeff_strongC} we find
that $C$ changes sign, from negative to positive, at $p^\ast =\frac{25}{27}=0.925925...$,
assuring the presence of CTPs for $p\in (p^\ast,1)$.
It should be noted that taking the limit $p\to 1$
in Eqs.~\eqref{eq:CTP_as_a_fnc_p_Delta},\eqref{eq:CTP_as_a_fnc_p_beta}
reproduces the known~\cite{Barre2001} result $(\Delta^\ast,\beta^\ast)=(\frac{1}{3}\ln 4,3)$.

CEPs satisfy the constraints of a first order transition $f(m_c,\beta_c) = f(0,\beta_c)$
and $f'(m_c,\beta_c) = f'(0,\beta_c)$, where $m_c$ is the nonzero critical magnetization, 
together with the continuous transition ($\lambda$ line) equation Eq.~\eqref{eq:can_2nd_strong}.
Since multicritical points are often very close to each other, CEPs may be detected using the truncated free energy Eq.~\eqref{eq:lada_expansion_strong}
where $A=0$ , leading to the condition~\cite{SumedhaMukherjee2020,MukherjeeSumedha2023,pmukherjee2021hase,Ishibashi}
\begin{equation}
\label{eq:CEP}
\frac{C^2}{4 B D} = 1\;,\ C<0\;.
\end{equation}
Eq.~\eqref{eq:CEP} may be particularly useful in probing the range of concentrations
where CEPs survive. To verify 
that $C$ is negative, it is useful to treat the expression in the large brackets of Eq.~\eqref{eq:landau_coeff_strongC} as a third order polynomial $p(v)$
defined on $(-\infty,\infty)$.
Now, since $p(v)\to \pm \infty$ as $v\to \pm \infty$ and $p(0) > 0$, $p(v)$
has a single negative root, assuring $p(v)>0$ and hence $C<0$ for positive $v$, if the discriminant of $p(v)$, proprietorial to $-14553 p^4+28080 p^3-13520 p^2$, is negative. The latter condition is satisfied for 
$p < p^{\ast\ast}=\tfrac{4680-52 \sqrt{15}}{4851} =0.923233...\ .$
It can be shown, using Wolfram Mathematica\textsuperscript{\textregistered},
that $D$ is positive for $p<p^{\ast\ast},\ \forall v>1$.
One can now deduce that 
room is left for CEPs at high concentrations.



\subsection{Microcanonical ensemble}
We now turn to the microcanonical ensemble, where the second-order line and the \textit{microcanonical tricritical point} (MTP) are found in a similar fashion to the canonical case. We start with expanding $s$ in small $m$ 
\begin{equation}
s = s_0 + am^2 + bm^4 + cm^6 +O(m^7)\;,
\end{equation}
where
\begin{equation}
\label{eq:mic_coeff_s0}
s_0 =
-\left(1-\frac{\epsilon}{\Delta p}\right)\ln \left(1-\frac{\epsilon}{\Delta p}\right)-\frac{\epsilon}{ \Delta p} \ln \left(\frac{\epsilon}{2 \Delta p}\right)\;,
\end{equation}
and
\begin{eqnarray}
\label{eq:mic_coeff_a_b}
a  & = &
\frac{\Delta p }{2\epsilon} + \frac{1}{2\Delta p}\ln\left(\frac{\epsilon}{2(\Delta p-\epsilon)}\right)\;,\nonumber\\
b &=&
-\frac{\Delta^3 p^3}{12 \epsilon^3}+\frac{\Delta p}{4 \epsilon^2}-\frac{1}{8 \Delta \epsilon p}+\frac{1}{8 \Delta p (\epsilon-\Delta p)}\;,\nonumber\\
c &=&-\frac{\Delta^5 p^5}{30 \epsilon^5}+\frac{\Delta^3 p^3}{8 \epsilon^4}-\frac{\Delta p}{8 \epsilon^3}+\frac{1}{48 \Delta \epsilon^2 p}\nonumber\\
&-&\frac{1}{48 \Delta p (\epsilon-\Delta p)^2}\;.
\end{eqnarray}
In order to obtain a temperature dependence of the coefficients $a,b,c$, the microcanonical definition of temperature, $\beta = \partial s/\partial\epsilon$, 
is applied to 
Eq.~\eqref{eq:mic_coeff_s0}.
This yields
\begin{equation}
\label{eq:epsilon_vs_beta}
\epsilon = \frac{2 \Delta p}{e^{\beta \Delta p}+2}\;,
\end{equation}
which, when substituted into Eq.~\eqref{eq:mic_coeff_a_b}, gives
\begin{eqnarray}
a &=& \frac{2(\beta-1)-e^{\beta\Delta p}}{4}
\label{eq:mic_coeff_temprrature_a}\;,\\
b &=& -\frac{e^{-\beta \Delta p} \left(e^{\beta \Delta p}+2\right)^2}{96 \Delta^2 p^2}
\label{eq:mic_coeff_temprrature_b}\;.\\
&\times &\Big(\Delta p e^{\beta \Delta p} \left(\Delta p \left(e^{\beta \Delta p}+2\right)-6\right)+6\Big)\nonumber \label{eq:mic_coeff_temprrature_c}\;.\\
c &=&
-\frac{(e^{\beta\Delta p}+2)^3}{1920 \Delta^3 p^3 e^{2\beta\Delta p}}\nonumber\\
&\times & \Big(2 \Delta^3 p^3 e^{4\beta\Delta p}+(8 \Delta^3 p^3-15 \Delta^2 p^2) e^{3\beta\Delta p}\nonumber\\
&+& (8\Delta^3 p^3-30 \Delta^2 p^2+30 \Delta p) e^{2\beta\Delta p}-10 e^{\beta\Delta p}+20\Big)\nonumber\\
\end{eqnarray}
Setting $a=0$, the second order line obeys
\begin{equation}
\label{eq:2nd_mic}
e^{\beta\Delta p} = 2(\beta -1)\;.
\end{equation}
By further equating $b$ to zero we have
\begin{equation}
\label{eq:MTP_mic}
6e^{-\beta\Delta p}=\Delta p\left(6-\Delta p \left(e^{\beta \Delta p}+2\right)\right)\;.
\end{equation}
One can observe that Eqs.~\eqref{eq:epsilon_vs_beta},\eqref{eq:2nd_mic},\eqref{eq:MTP_mic}
can be realized as functions of $\beta $ and the products $\Delta p$.
It is thus convenient to use the substitution $K = \Delta p$ to, e.g., relate the critical energy to the critical temperature by plugging Eq.~\eqref{eq:epsilon_vs_beta} into Eq.~\eqref{eq:2nd_mic}. This gives
\begin{equation}
\label{eq:epsilon_c}
\epsilon_c = \frac{K}{\beta(K)}\;,
\end{equation}
where $\beta(K)$ is now regarded as a $K$-dependent solution of Eq.~\eqref{eq:2nd_mic}.

Interestingly, the quantities  $K^\ast= \Delta^\ast p,\beta^\ast$,
where $\Delta^\ast,\beta^\ast$, constituting the MTP by simultaneously solving Eqs.~\eqref{eq:2nd_mic},\eqref{eq:MTP_mic}, do not vary with $p$. To see this it 
is useful to consider the system 
\begin{equation}
\label{eq:flow}
\bm F(\bm x)=0\;,
\end{equation}
where $\bm x =(K,\beta)$ and $\bm F(\bm x)$ is an auxiliary vector field
whose components are determined by Eqs.~\eqref{eq:2nd_mic},\eqref{eq:MTP_mic}.
It is shown in Appendix \ref{eq:AppC} that Eq.~\eqref{eq:flow}
has exactly two solutions. These are found to be $\bm x_1^\ast=(0.4624....,3.0271....)$ and
$\bm x_2^\ast=(0.40955....,5.1915....)$. 
The solution $\bm x_1^\ast$
agrees with the MTP of the homogeneous model ($p=1$)~\citep{Barre2001}. This makes $\bm x_1^\ast$  a physical solution leading to
\begin{eqnarray}
\label{eq:physical_sol_K_beta}
\Delta^\ast &=& \frac{K^\ast}{p}\;,K^\ast=0.4624....\;,\nonumber\\
\beta^\ast &=& 3.0271....\;.
\end{eqnarray}

\begin{figure}[ht]
\centering
\includegraphics[width=1\columnwidth]{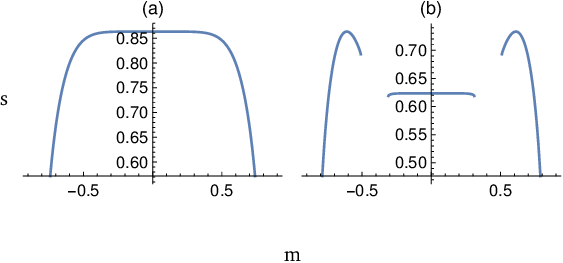}
\caption{Entropy versus magnetization in the invariant form for $p=0.5$, associated
with each of the two solutions of system~Eq.~\eqref{eq:flow}. Panel (a) shows
the physical solution $\bm{x}_1^\ast$, with energy $\epsilon_1^\ast=
0.1527....$, and panel (b) shows the unphysical solution $\bm{x}_2^\ast$, with
energy $\epsilon_2^\ast= 0.0788....~$.
}
\label{fig:Fig1}
\end{figure}

It can be verified that inserting $\bm x_1^\ast$ into Eq.~\eqref{eq:mic_coeff_temprrature_c}
makes the sixth order coefficient negative.
This guarantees the existence of MTPs at any $p$. One can also note that 
Eq.~\eqref{eq:epsilon_c} is satisfied, in particular, at the MTP, hence $K,\beta(K)$ can be replaced with $K^\ast,\beta^\ast$ to produce $\epsilon^\ast = K^\ast/\beta^\ast=0.1527...$ which is another constant quantity. Then, plugging $\epsilon^\ast,K^\ast$ into Eqs.~\eqref{eq:normalized},\eqref{eq:entropy} 
implies that the entropy at the MTP, $s^\ast$, is essentially a function of $m$. Consequently, the graph of $s^\ast(m)$ is invariant under changes in the concentration $p$. The solution $\bm x_2^\ast$ also results in an invariant entropy graph. The two scenarios
are depicted in Fig.~\ref{fig:Fig1} for (arbitrarily chosen) $p=0.5$.
Fig.~\ref{fig:Fig1}(a) shows the invariant graph at the MTP and in Fig.~\ref{fig:Fig1}(b),
the associated solution is $\bm x_2^\ast$. Though the entropy curve has a local broad maximum
at $m=0$, it is globally maximized by nonzero magnetizations. Furthermore, 
domains of magnetizations that are inaccessible to the system  decompose the entropy curve
into three disjoint segments. The fact that the entropy is not globally maximized by the zero magnetization rules out $\bm x_2^\ast$ from being a physical solution. This 
guaranties the MTP Eq.~\eqref{eq:physical_sol_K_beta} is unique.

As expected, the canonical second order line Eq.~\eqref{eq:can_2nd_strong}
and the microcanonical line Eq.~\eqref{eq:2nd_mic} coincide in the marginal case where
$p=1$.
In particular, the picture within the two ensembles when $p=1$,
is in agreement with previous studies~\cite{schreiber2023ensemble,Barre2001}, where the 
identical canonical and microcanonical lines terminate at the CTP $(\frac{1}{3}\ln 4,3)$ and the MTP due to Eqs.~\eqref{eq:2nd_mic},\eqref{eq:MTP_mic}, respectively.
It should be also noted that the Ising-like, $\Delta$-independent, critical point $\beta_c = \frac{3}{2}$ is commonly approached as $p\to 0$.

\subsection{Critical behaviour}
Here, we study the critical behaviour of the model 
by analysing the canonical and microcanonical phase diagrams 
in CF-temperature parameter space.
The qualitative behaviour of the thermodynamic potentials 
as functions of the magnetization,
is determined by the parameters chosen from the associated phase diagrams.
Several related plots are provided.

\begin{figure}[ht]
\centering
\hspace*{-0.5cm}
\includegraphics[width=1.15\columnwidth]{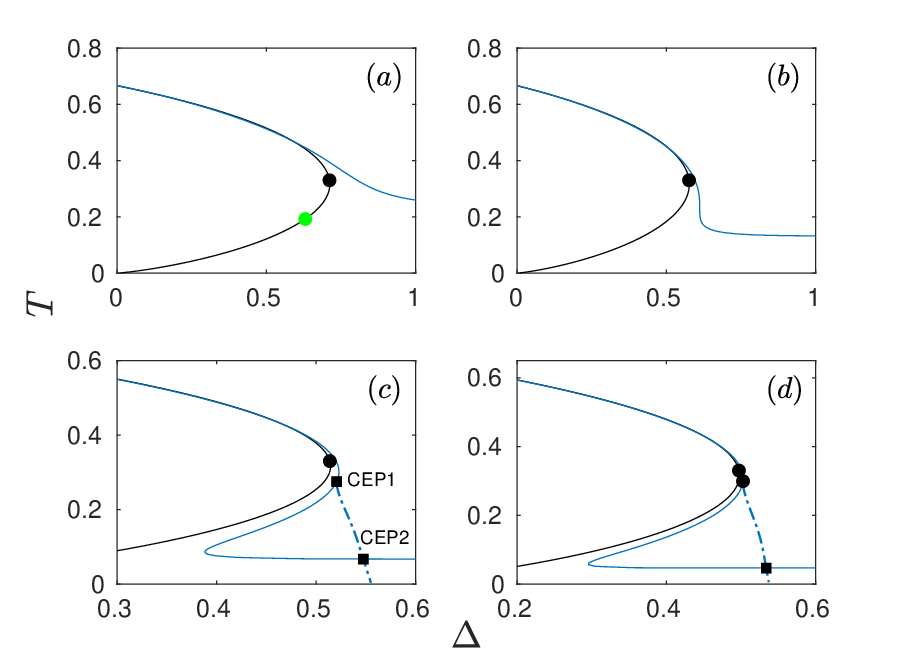}
\caption{Canonical (blue) and microcanonical (black) phase diagrams, together with
the multicritical points, in the $(\Delta,T)$ plane for increasing $p$
(decreasing disorder): (a) $p=0.65$; (b) $p=0.8029$; (c) $p=0.9$;
(d) $p=0.93$. Solid lines represent continuous transitions, and
dash-dotted lines mark discontinuous transitions. CTPs and MTPs are both
denoted by black solid circles, and CEPs by solid squares. The green solid
circle in (a) denotes the unphysical point $\bm{x}_2^\ast$.
}
\label{fig:Fig2}
\end{figure}

We start with a plot of the phase diagram for different concentrations, shown in Fig.~\ref{fig:Fig2}. 
Solid lines indicate continuous transitions while dash-dotted lines represent
discontinuous transitions. Tricritical points are denoted by solid circles while solid squares 
describe CEPs. The green solid circle in Fig.~\ref{fig:Fig2}(a) corresponds to the
unphysical point $\bm{x}_2^\ast$ and is therefore omitted from the
other panels. If the black curves are divided by the MTPs into lower and upper
branches, then, by continuity, the lower branches are unphysical
in all cases. As previously argued, MTPs are expected at any concentration, and indeed they appear in all four panels. CTPs, however, are absent in Figs.~\ref{fig:Fig2}(a),\ref{fig:Fig2}(b) 
and \ref{fig:Fig2}(c), where $p$ is too low. Although the emergence of canonical CEPs cannot be ruled out for the concentrations in Figs.~\ref{fig:Fig2}(a) and \ref{fig:Fig2}(b), we could
not locate them numerically. Accordingly, the discontinuous transition
line that would meet such CEPs does not appear, and its existence
in this range remains questionable.
In Fig.~\ref{fig:Fig2}(c) two CEPs (denoted by CEP1, CEP2) are observed. Consequently, the
transition pattern is broken into a continuous transition terminating at CEP1,  
a discontinuous transition connecting CEP1 and CEP2, and another continuous transition
originating at CEP2. A similar pattern can be seen in Fig.~\ref{fig:Fig2}(d) where
$p$ is high enough such that a CTP arises, seemingly, instead of the upper CEP
in Fig.~\ref{fig:Fig2}(c).

\begin{figure}[htbp]
    \centering
    \begin{subfigure}[b]{0.45\textwidth}
        \centering
        \includegraphics[width=\columnwidth]{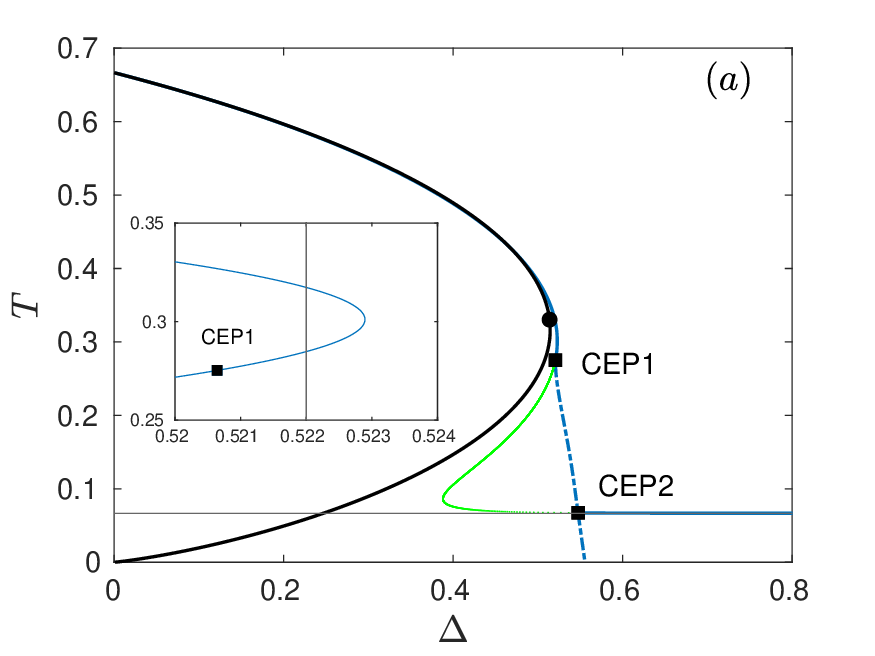}
        \label{fig:top}
    \end{subfigure}
    
    \vspace{-0.5cm} 
    
    \begin{subfigure}[b]{0.45\textwidth}
        \centering
        \includegraphics[width=\columnwidth]{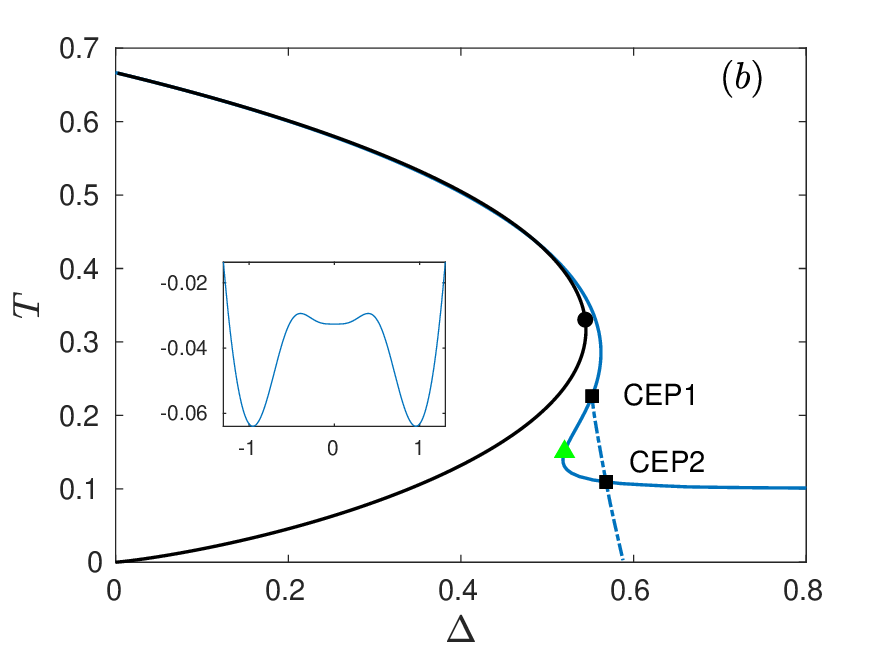}
        \label{fig:bottom}
    \end{subfigure}
    
    \caption{Canonical (blue) and microcanonical (black) phase diagrams for different
concentrations. The blue curves give the complete canonical transition
picture, which consists of three parts: a continuous (solid) line that
ends at CEP1, a discontinuous (dash-dotted) line, and another continuous
line originating at CEP2. Solid circles denote MTPs, and solid squares
mark the canonical CEPs.
(a) $p=0.9$. The green dotted line denotes the unphysical part from
Fig.~\ref{fig:Fig2}(c) that connects CEP1 and CEP2. The horizontal solid
line shows the asymptotic behaviour given by Eq.~\eqref{eq:T_inf}. The
inset shows a close-up of the neighbourhood of CEP1. Reentrance occurs
along the vertical line $\Delta = 0.522$.
(b) $p=0.85$. The inset shows the free energy at the point indicated by
the green solid triangle.} 
    \label{fig:Fig3}
\end{figure}

In Fig.~\ref{fig:Fig3} we delve deeper into the canonical phase diagram
of Fig.~\ref{fig:Fig2}(c). For instance, we note that on the curve connecting
CEP1 and CEP2 in Fig.~\ref{fig:Fig3}(a) (denoted by a green dotted line), the fourth order term $B$ (Eq. \ref{eq:landau_coeff_strongB}) is negative.
This rules out a continuous transition and makes this curve unphysical.
It should be noted, however, that $B>0$ is a \textit{necessary}
condition for a continuous transition. Fig.~\ref{fig:Fig3}(b) exemplifies its insufficiency. We take $p=0.85$ and choose an arbitrary point (denoted by a solid triangle) along the curve determined by $A=0$, that connects CEP1 and CEP2. Indeed, we have $B>0$ at that point and a
local broad minimum is attained. Still, the free energy is globally minimized, as realized from the inset, by nonzero magnetizations. One can thus presume the entire path between the two CEPs does not describe a continuous transition, that is, also unphysical.
Both panels suggest that, in general, the canonical phase diagram consists
of a continuous transition branch that terminates at either a CTP or a CEP.
A discontinuous transition line originates at this multicritical point and
ends at a second CEP, where it meets another continuous transition branch.
This branch should extend to 
infinity and approach, according to Eq.~\eqref{eq:can_2nd_strong},  
\begin{equation}
\label{eq:T_inf}
T= \frac{2}{3}(1-p)\;,
\end{equation}
as can be plausibly realized from Fig.~\ref{fig:Fig3}(a)
and in agreement with~\cite{Benyoussef1987}. Note that for both values of $p$ the points CEP1 
are located below the turning point of the $\lambda$ line. This means that the canonical phase diagram has a re-entrant part where the system goes from the paramagnetic phase to the ferromagnetic phase and then to the paramagnetic phase again. This can be clearly realized from the inset
of Fig.~\ref{fig:Fig3}(a) where a typical path along which re-entrant occurs is sketched.
In contrast, in the the microcanonical ensemble $\beta^\ast$ is fixed, and the phase boundary is monotonic (see, e.g, the black physical upper branch in Fig.~\ref{fig:Fig3}(b)). Consequently, no re-entrant phenomenon can be observed.

\begin{figure}[ht]
\centering
\includegraphics[width=\columnwidth]{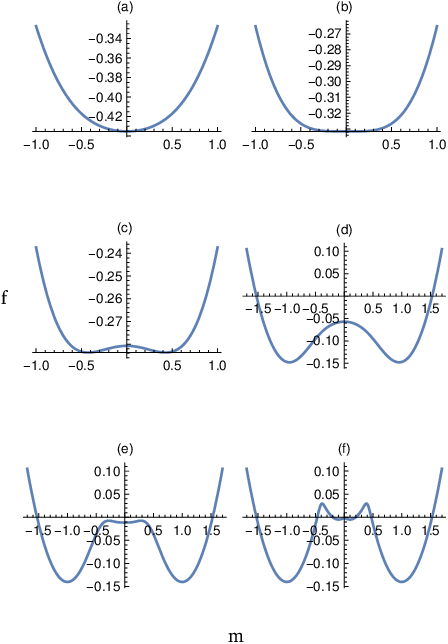}
\caption{Free energy as a function of magnetization for $p=0.9$, $\Delta=0.4$, and
decreasing temperatures: (a) $T=0.5884$; (b) $T_c=0.4884$, satisfying Eq.~\eqref{eq:can_2nd_strong}; (c) $T=0.4384$; (d) $T=0.1884$;
(e) $T=0.0884$, where the free energy has a local minimum at $m=0$;
(f) $T=0.0184$, where two local minima appear at $m\neq 0$.
}
\label{fig:Fig4}
\end{figure}

Variation of the free energy with the magnetization
is displayed in Fig.~\ref{fig:Fig4} for a few temperatures, keeping $p$ and $\Delta$ fixed.
Fig.~\ref{fig:Fig4}(a) and Fig.~\ref{fig:Fig4}(b)
describe the system above and at $T_c$, respectively.
At lower temperatures the typical second order double well appears
(see Fig.~\ref{fig:Fig4}(c) and Fig.~\ref{fig:Fig4}(d)). One can see in Fig.~\ref{fig:Fig4}(e) that a local minimum appears at zero magnetization as the temperature is further decreased further.
At an even lower temperature, the local minimum splits into two local
minima at nonzero magnetizations, as can be seen in Fig.~\ref{fig:Fig4}(f).
It can be verified that these minima are smeared out in the homogeneous
model. One can thus infer that the inert spins enable metastable
long-range order.

\begin{figure}[ht]
\centering
\includegraphics[width=\columnwidth]{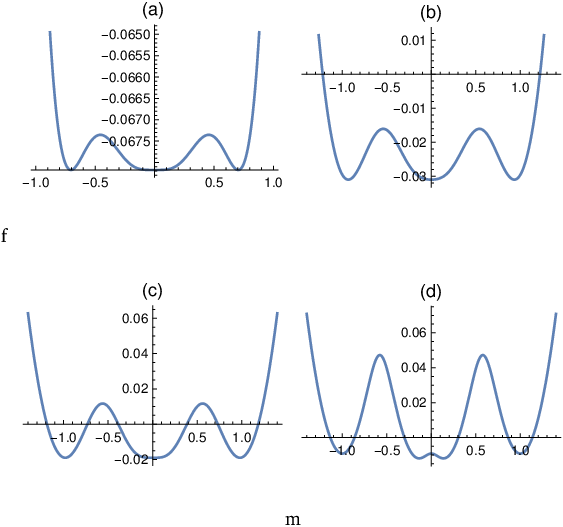}
\caption{Free energy as a function of magnetization at several points $(\Delta,T)$
along the dash-dotted line in Fig.~\ref{fig:Fig3}(b) ($p=0.85$):
(a) $\mathrm{CEP1}=(0.5519,0.2262)$, where narrow minima at $m\neq 0$
coexist with a broad minimum at $m=0$; (b) $(0.5616,0.1511)$;
(c) $\mathrm{CEP2}=(0.5681,0.1093)$, qualitatively similar to (a);
(d) $(0.5774,0.0558)$. Note that in (d) the free energy has a local
maximum at $m=0$ and hence does not describe a discontinuous transition.
}
\label{fig:Fig5}
\end{figure}

In Fig.~\ref{fig:Fig5} we present plots of the free energy 
associated with the first order line of Fig.~\ref{fig:Fig3}(b), against the magnetization. 
Fig.~\ref{fig:Fig5}(a) where the line originates at 
CEP1 = $(0.5519,0.2262)$, Fig.~\ref{fig:Fig5}(b) at the point $(0.5616,0.1511)$ along the line, 
Fig.~\ref{fig:Fig5}(c) CEP2 = $(0.5681,0.1093)$ where the first order line meets the second order line again. In all panels the expected coexistence of disordered (ordered) states
associated with zero (nonzero) magnetization is observed. In Fig.~\ref{fig:Fig5}(d), however, the free energy has a local maximum at $m=0$ hence the coexistence is violated. This confirms that the first order line terminates at CEP2, making the dash-dotted part below CEP2 in Fig.~\ref{fig:Fig3}(b) (and presumably in Fig.~\ref{fig:Fig3}(a)) unphysical.

\begin{figure}[ht]
\centering
\includegraphics[width=\columnwidth]{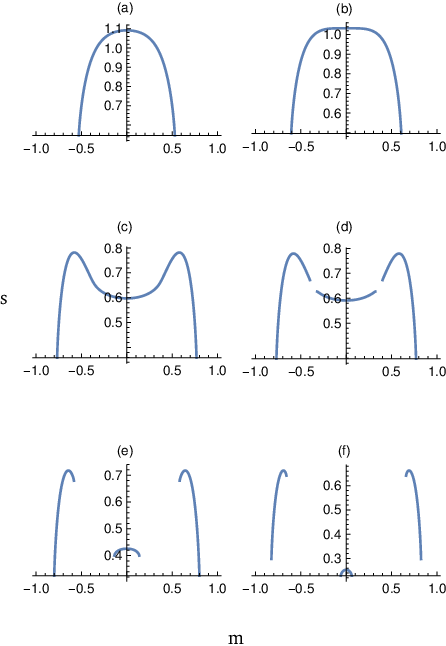}
\caption{Entropy as a function of magnetization for $p=0.9$, $\Delta=0.4$, and
decreasing energies: (a) $\epsilon=0.22$; (b) the critical energy
$\epsilon_c = 0.1761$..., given by Eq.~\eqref{eq:epsilon_c};
(c) $\epsilon=0.065$; (d) $\epsilon=0.064$; (e) $\epsilon=0.04$;
(f) $\epsilon=0.02$.}
\label{fig:Fig6}
\end{figure}

For a qualitative comparison with the canonical picture, we plot in Fig.~\ref{fig:Fig6} the entropy versus the magnetization for decreasing values of the energy and the same parameters $p=0.9,\Delta=0.4$ of Fig.~\ref{fig:Fig4}.
Fig.~\ref{fig:Fig6}(a) describes a totally disordered system. In Fig.~\ref{fig:Fig6}(b)
the system at criticality with energy $\epsilon_c$ (see Eq.~\eqref{eq:epsilon_c}).
In Fig.~\ref{fig:Fig6}(c) the energy is below $\epsilon_c$ and the entropy is now maximized
by the ordered states. The energy is further lowered in Fig.~\ref{fig:Fig6}(d)
and voids in the entropy graph are initially seen. These voids expand in FIG \ref{fig:Fig6}(e)
and \ref{fig:Fig6}(f) and the entropies acquire a local maximum at $m=0$.

\section{Concluding remarks}
\label{sec:conclusion}
A system with LRI obeying the BEG Hamiltonian has been studied.
It has been found that the canonical and microcanonical ensembles
result in completely different phase diagrams in $(\Delta,T)$ plane.
In the microcanonical ensemble, the system exhibits a continuous transition
within a finite range of CF values for every $p \in (0,1)$. 
In the canonical ensemble, 
for every $p \in (0,1)$, the ferromagnetic phase is stable for arbitrarily large $\Delta$. In fact, dilution of the
homogeneous model by any finite fraction $\varepsilon$ of inert spins
effectively generates an independent subsystem that is responsible for
this stability.

This special role of the inert spins can be heuristically demonstrated 
by noting that for sufficiently large $\Delta$, 
the active spins typically occupy the zero state at any temperature.  
This results in a temperature independent constant (zero) free energy.
Furthermore, the total energy due to LRI between active and inert spins is equal, on average, to zero.
Consequently, the entire system benefits from the presence of the inert
spins. More precisely, the system can lower its free energy by forming a
``large'' component of size $\varepsilon N$ consisting solely of these spins. The total change in the free energy is then given by
$\Delta F = -\varepsilon^2 N/2 + T\varepsilon N \ln 3$. Clearly, $\Delta F$
is negative for $T<\tilde{T}_c$, where
$\tilde{T}_c = (1-p)/\ln 9 \approx 0.455\,(1-p)$
(cf.~Eq.~\eqref{eq:T_inf}) is the pseudocritical temperature obtained by
setting $\Delta F=0$ with $\varepsilon$ replaced by $1-p$. This concludes
the argument.

Another interesting observation is the emergence of canonical CEPs, apart from tricritical points which coexist within the canonical ensemble as well as the microcanonical ensemble.
This observation has been made in part~\cite{Benyoussef1987}, that is, unlike in this reference
where CEPs exist for $\frac{8}{9}\leq  p < \frac{25}{27}$, our findings suggest that CEPs survive for a broader range of concentrations (see, e.g, Fig.~\ref{fig:Fig3}(a) where CEP1 and CEP2 are 
detected for $p=0.85$). 	

We also point out the markedly different dependence of tricritical points
on $p$ in the two ensembles. Both coordinates of the CTPs vary with $p$. Furthermore, since $v^\ast$ in
Eq.~\eqref{eq:CTP_as_a_fnc_p_Delta} is bounded, the region occupied by the
CTPs is also bounded.
MTPs, on the other hand, originate from the constant quantities in
Eq.~\eqref{eq:physical_sol_K_beta} and therefore have a fixed coordinate,
$\beta^\ast$. The other coordinate, $\Delta^\ast = K^\ast/p$, is finite for
every $p \in (0,1)$ and grows without bound only as $p \to 0$. Hence, MTPs
never disappear; they merely shift to larger $\Delta$ as $p$ decreases.

To conclude, we note that the distortion of the phase diagrams of
homogeneous models by the introduction of impurities is a rather general
phenomenon~\cite{aizenman1989} that occurs even in models with short-range
interactions~\cite{Schreiber_2022}.


\section{Appendix}
Some equations which appear in the main text are derived in detail in the present section.
\appendix
\section{Free energy, random fields}
\label{AppA}
We derive here Eqs.~\eqref{eq:free_energy_Ns1},\eqref{eq:mag}.
To this end we consider $n$ replicas of the system described by the Hamiltonian Eq.~\eqref{eq:H} and write
\begin{eqnarray}
\label{eq:Z^n_app}
Z^n  &=& \prod_\alpha \Tr_{\alpha}e^{\frac{\beta}{2N}\left(\sum_i\sigma_{i\alpha}\right)^2-\beta\sum_i\Delta_{i\alpha}\sigma_{i\alpha}^2}\;,
\end{eqnarray}
where $\alpha=1,...,n$ and $\Tr_\alpha$ represents the trace over all configurations of replica $\alpha$.
Using the integral identity 
\begin{equation}
\label{eq:gauss}
e^{\frac{\nu ^2}{2\mu}} = \sqrt{\frac{\mu}{2\pi}}\int_{-\infty}^\infty dx e^{-\frac{1}{2}\mu x^2+\nu x}\;,\end{equation}
with $\mu=N\beta$ and $\nu = \beta\sum_i\sigma_{i\alpha}$, 
Eq.~\eqref{eq:Z^n_app} takes the form
\begin{eqnarray}
\label{eq:Z_alpha_replica}
Z^n
&=&\prod_\alpha \sqrt{\frac{N\beta}{2\pi}}\int_{-\infty}^\infty dx_\alpha
e^{-N\frac{1}{2} \beta x_\alpha^2}\nonumber\\
&\times &\Tr_\alpha e^{\beta x_\alpha\sum_i\sigma_{i\alpha} -\beta\sum_i
\Delta_{i\alpha}\sigma_{i\alpha}^2}\;.
\end{eqnarray}
Since in each replica the single particle fields are \textit{i.i.d},
we can further write

\begin{eqnarray}
\label{eq:Z_alpha_app}
Z^n
&=&\left(\frac{N\beta}{2\pi}\right)^{n/2}\int_{-\infty}^\infty dx_1\int_{-\infty}^\infty dx_2...\int_{-\infty}^\infty dx_n \nonumber\\
&\times &\exp {\left[-N\left(\frac{1}{2}\beta\sum_\alpha x_\alpha^2-\ln \Tr_{\sigma_\alpha} e^{-\beta H_r}\right)\right]}\;,\nonumber\\
\end{eqnarray}
where $\Tr_{\sigma_\alpha}$ stands for trace over states of a single spin in replica $\alpha$, 
subject to a random field $\Delta_\alpha/\Delta\sim \mathrm{Ber}(p)$,
and $H_r$ is the \textit{replica Hamiltonian}
\begin{equation}
\label{eq:H_r}
H_r = -\sum_\alpha x_\alpha\sigma_\alpha  + \sum_\alpha \Delta_\alpha\sigma_\alpha^2\;.
\end{equation}
We now implement the saddle point approximation to calculate Eq.~\eqref{eq:Z_alpha_app}.
To this end, we first
minimize the expression in the regular brackets in Eq.~\eqref{eq:Z_alpha_app}
with respect to $x_\gamma,\ \gamma=1,...,n$. The minimizers, $m_\gamma$, satisfy
\begin{equation}
\label{eq:sigma_av}
m_\gamma = \langle\sigma_\gamma\rangle_{H_r}\;,
\end{equation}
where $\langle...\rangle_{H_r}$ is the thermal average with respect to ${H_{r}}$.
Next, we assume the \textit{replica symmetry ansatz} $m_\gamma = m$.
This allows to uniformly 
maximize the exponential in Eq.~\eqref{eq:Z_alpha_app} and write
\begin{eqnarray}
\label{eq:Z_uniform}
Z^n & = & \exp \Big\{-n\Big[N\nonumber\\
&\times &\Big(\frac{1}{2}\beta m^2-\ln \Tr_{\sigma} e^{\beta m \sigma-\beta\Delta_\sigma\sigma^2}\Big)+o(N)\Big]\Big\}\;,\nonumber\\
\end{eqnarray}
where, by homogeneity, the index $\alpha$ is omitted from the spin
variables and the random field is denoted now by $\Delta_\sigma$. Finally, we apply the identity  $\ln Z = \lim_{n\to 0} (Z^n-1)/n$
to Eq.~\eqref{eq:Z_uniform} and take the thermodynamic limit
of the disorder average
$\lim_{N\to\infty}-\frac{1}{N}\langle\ln Z\rangle$. This leads to
\begin{eqnarray}
\label{eq:Z_anstaz}
\beta f 
 &=&
\frac{1}{2}\beta m^2-
\left\langle\ln \Tr_{\sigma} e^{\beta m \sigma-\beta\Delta_\sigma\sigma^2}\right\rangle\;.
\end{eqnarray}
Equation Eq.~\eqref{eq:free_energy_Ns1} now immediately follows from Eq.~\eqref{eq:Z_anstaz}.

By assuming the quantities $m_\gamma$ are narrowly distributed around 
their disorder average, Eq.~\eqref{eq:sigma_av} yields Eq.~\eqref{eq:mag}.
Strictly speaking, take
\begin{equation} 
\label{eq:sigma_av_dis}
\langle m_\gamma\rangle = p\langle\sigma_\gamma\rangle_{H_r^\Delta}
+(1-p)\langle\sigma_\gamma\rangle_{H_r^0}\;,
\end{equation}
where $H_r^\Delta,H_r^0$ denote the replica Hamiltonian
Eq.~\eqref{eq:H_r} at $\Delta_\alpha = \Delta,\Delta_\alpha = 0$, respectively.
Then replace $\langle m_\gamma\rangle$ with $m_\gamma = m$.

\section{Free energy, random spins}
\label{AppB}
Eqs.~\eqref{eq:Z_strong_main},\eqref{eq:free_energy_Ns} are derived in the present appendix.
We start with the partition function of the system 
governed by the Hamiltonian Eq.~\eqref{eq:H_new}
and employ Eq.~\eqref{eq:gauss} 
with $\mu=N\beta$ and $\nu = \beta\sum_i\sigma_i$, to write
\begin{eqnarray}
\label{eq:Z_main_text}
Z_a &=& \Tr_{\{\sigma\}} e^{\frac{\beta}{2N}\left(\sum_i\sigma_i\right)^2-\beta \Delta\sum_{I}\sigma_{I}^2}\nonumber\\
&=&\sqrt{\frac{N\beta}{2\pi}} \int_{-\infty}^\infty dx
e^{-\frac{1}{2}N\beta x^2}\Tr_{\{\sigma\}}e^{\beta x \sum_i \sigma_i-\beta\Delta\sum_I\sigma_I^2}\;,
\nonumber\\
\end{eqnarray}
where the trace is over all spin configurations $\{\sigma\}$.
The trace can be decomposed into
trace over configurations of active spins $\{\sigma_a\}$
and inert spins $\{\{\sigma\}\setminus\{\sigma_a\}\}$, namely,

\begin{eqnarray}
\label{eq:Z_Ns}
Z_a &=&\sqrt{\frac{N\beta}{2\pi}} \int_{-\infty}^\infty dx 
e^{-\frac{1}{2}N\beta x^2}\nonumber\\
&\times &
\Tr_{\{\sigma_a\}}e^{\beta x \sum_I \sigma_I-\beta\Delta\sum_I\sigma_I^2}
\Tr_{\{\{\sigma\}\setminus\{\sigma_a\}\}}e^{\beta x \sum_i \sigma_i^2}\nonumber\\
&=&\sqrt{\frac{N\beta}{2\pi}} \int_{-\infty}^\infty dx
e^{-\frac{1}{2}N\beta x^2}\nonumber\\
&\times &
\left(1+2e^{-\beta\Delta}\cosh(\beta x)\right)^{N_a}
\left(1+2\cosh(\beta x)\right)^{N-N_a}\nonumber\;.\\
\end{eqnarray}
Finally, exponentiating the last row of Eq.~\eqref{eq:Z_Ns} leads to the desired equations.
\section{Solutions to $\bm F(\bm x)=0$}
\label{eq:AppC}
In the present appendix it is shown that the system $\bm F(\bm x)=0$ where $\bm x = (K,\beta)$
and $\bm F(\bm x)$ is a vector field with components
\begin{eqnarray}
F_1(\bm x) &=&e^{\beta K}-2 (\beta-1)\label{eq:system1}\;,\\
F_2(\bm x) &=& 6 e^{-\beta K} - 6K + K^2 \left(e^{\beta K}+2\right)\label{eq:system2}\;,
\end{eqnarray}
has exactly two solutions.

We start with plugging $e^{\beta K} = 2(\beta-1)$ 
into Eq.~\eqref{eq:system2} and solving $F_2(\bm x)=0$ for $K$ which gives
\begin{equation}
\label{eq:K_pm}
K_\pm = \frac{3}{2\beta}\left(1\pm\sqrt{\frac{\beta-3}{3\beta-3}}\right)\;.
\end{equation}
Next, Eq.~\eqref{eq:K_pm} is substituted into Eq.~\eqref{eq:system1}
and a careful look is taken at the equation $F_1(K_\pm,\beta)=0$.
We note that $e^{\beta K_\pm}$ are greater than $2\beta -2$ at $\beta  =3$ ($\sqrt{e^3}>4$). It can also be easily verified that $e^{\beta K_+}$ is a bounded concave function in $I=[3,\infty)$. These imply 
$e^{\beta K_+}$ crosses $2\beta -2$ exactly once.
As for the other branch, $e^{\beta K_-}$, 
one can check it is monotonically decreasing in $I$, hence also crosses $2\beta-2$ exactly once.
In other words, each branch produces a single solution 
of Eq.~\eqref{eq:system1} in $I$. 
Denote these solutions by $\beta^\ast_\pm$. Then, direct substitution 
of each solution into the proper branch gives 
two distinct values $K^\ast_\pm$. The system 
therefore has exactly two solutions $(K^\ast_+,\beta^\ast_+),(K^\ast_-,\beta^\ast_-)$. 
The physical solution employed in Eq.~\eqref{eq:physical_sol_K_beta} is nothing but $(K^\ast_-,\beta^\ast_-)$.

\sloppy
\bibliography{references_random.bib}

\end{document}